\documentclass[fleqn,usenatbib]{mnras}
\usepackage{mathptmx}
\usepackage{soul}
\usepackage[T1]{fontenc}
\usepackage{ae,aecompl}
\usepackage{graphicx}	
\usepackage{amsmath}	
\usepackage{amssymb}	

\usepackage{standalone}
\usepackage{tikz}

\title[Real-Time Imaging on the LWA]{A Real-Time, All-Sky, High Time Resolution, Direct Imager for the Long Wavelength Array} 
\author[J. Kent et al.]{
James Kent,$^{1}$\thanks{E-mail: jck42@cam.ac.uk}
Jayce Dowell,$^{2}$
Adam Beardsley,$^{3}$
\newauthor
Nithyanandan Thyagarajan,$^{4,3}$\thanks{Nithyanandan Thyagarajan is a Jansky Fellow of the National Radio Astronomy Observatory.}
Greg Taylor$^{2}$
and Judd Bowman$^{3}$
\\
$^{1}$Cavendish Laboratory, University of Cambridge, UK\\
$^{2}$Department of Physics and Astronomy, University of New Mexico, Albuquerque, NM, USA\\
$^{3}$School of Earth and Space Exploration, Arizona State University, Tempe, AZ, USA\\
$^{4}$National Radio Astronomy Observatory, Socorro, NM, USA
}

\date{Accepted XXX. Received YYY; in original form ZZZ}

\pubyear{2018}

\begin{document}
\label{firstpage}
\pagerange{\pageref{firstpage}--\pageref{lastpage}}
\maketitle

\begin{abstract}
    
The future of radio astronomy will require instruments with large collecting areas for higher sensitivity, wide fields of view for faster survey speeds, and efficient computing and data rates relative to current capabilities.  We describe the first successful deployment of the E-field Parallel Imaging Correlator (EPIC) on the LWA station in Sevilleta, New Mexico, USA (LWA-SV). EPIC is a solution to the computational problem of large interferometers. By gridding and spatially Fourier transforming channelised electric fields from the antennas in real-time, EPIC removes the explicit cross multiplication of all pairs of antenna voltages to synthesize an aperture, reducing the computational scaling from $\mathcal{O}(n_a^2)$ to $\mathcal{O}(n_g \log_2 n_g)$, where $n_a$ is the number of antennas and $n_g$ is the number of grid points. Not only does this save computational costs for dense arrays but it produces very high time resolution images in real time. The GPU-based implementation uses existing LWA-SV hardware and the high performance streaming framework, Bifrost. We examine the practical details of the EPIC deployment and verify the imaging performance by detecting a meteor impact on the atmosphere using continuous all-sky imaging at 50 ms time resolution.

\end{abstract}

\begin{keywords}
instrumentation: interferometers -- radio continuum: transients -- telescopes
\end{keywords}

\section{Introduction}

Radio astronomy has been undergoing a renaissance in recent years,
with a number of new instruments, both built and in the proposal
and design phases. Future  instruments such as the Square Kilometre
Array \citep[SKA;][]{ska}, and current instruments such as the Long Wavelength Array \citep[LWA;][]{taylor_first_2012}, Canadian Hydrogen Intensity Mapping Experiment \citep[CHIME;][]{chime_frb} 
and the Hydrogen Epoch of Reionisation Array \citep[HERA;][]{hera}, are looking at using high density interferometric arrays with hundreds or thousands of individual antennas to facilitate wide-field, high sensitivity and angular resolution imaging of the sky.

There has also been a renewed focus on observations of transient
phenomena such as Fast Radio Bursts (FRBs), where the origins and physical mechanisms are an active area of research. Therefore the capability to detect and image these in real-time is of key scientific importance.  Interferometric measurements of FRBs have been previously achieved \citep{caleb_first_2017}, including by CHIME \citep{amiri_observations_2019}. High time resolution imaging of such phenomena would provide a significant new capability, by allowing dragnet surveys of the sky with wide field of view instruments.

Together, these two developments present a significant computational challenge for future interferometers, especially for the correlator.  The standard FX correlator, where the signal from each antenna is multiplied with the signals from every other antenna to produce ``visibilities'', mathematically defined as an outer product, scales as $\mathcal{O}(n_a^2)$, where $n_a$ is the number of antennas \citep{romney_lecture_1985}.  This scaling becomes problematic as proposed arrays will contain thousands of dipole elements.  All $n_a^2$ visibilities must be generated at the time resolution desired and subsequently gridded and then Fourier transformed to produce images, typically creating a bottleneck for high-time resolution studies. 

For some array geometries, the number of visibilities calculated can be reduced by omitting short baselines with little impact on point source imaging performance.  Fast convolution algorithms may also be used for correlation \citep{bunton_antenna_2011} to further reduce the computational costs to $\mathcal{O}(n_a^{3/2})$.

An alternative to full-field imaging with an FX correlator is to use a beamformer that provides the telescope's response to only a few chosen locations on the sky by summing over the voltages from all antennas with appropriate delays to direct the response in a particular direction.  The computational costs of a beamformer generally scale as $\mathcal{O}(n_a)$ per calculated beam and the output data volume is proportional only to the number of beams calculated.  This avoids the challenges of full-field imaging with an FX correlator, but with an associated compromise of limited sky coverage.

Direct Fourier transform imagers \citep{daishido_direct_1991,foster_et_al_2014} provide another alternative to both of the above approaches. Direct imaging forgoes the calculation of antenna cross products.  Instead, the antenna electric fields are gridded directly onto an aperture plane and Fourier transformed into an image plane. These images can be accumulated for noise reduction, in the same way visibilities are accumulated in FX correlators.

Theoretically they can provide significant potential scaling improvement by scaling as $\mathcal{O}(n_g\log{n_g})$ where $n_g$ is the number of grid points in the aperture, yielding a significant potential scaling advantage for high-density arrays \citep{morales_enabling_2011, thyagarajan_generic_2017}. Direct imagers facilitate full-field imaging at a high time resolution also because the output data volume can be much lower than for an FX correlator, scaling only as $n_g \approx n_a$ for a dense array. 

Previous direct imagers such as \cite{daishido_direct_1991,foster_et_al_2014} have relied on antennas being on a regular grid, which
limits their application from a scientific standpoint. For example, their uniform layouts yield point spread functions that  contain periodic grating responses that are not ideal for imaging applications. Further inherent assumptions about identical behaviour of antenna elements have to be made. As well as this, calibration still relies on using cross-correlated data products. \cite{morales_enabling_2011} proposed the MOFF formalism as a flexible generalization of the direct imaging approach. A framework is described which exploits the computational advantages provided by direct fourier transform imaging but with no limitations placed on the mixture of antenna elements or their placement, as well as producing fully calibrated images. In addition, provision is made for adaptive fourier optics which can correct for non-coplanar array effects as well as antenna dependent terms. Visibilities from an FX correlator can be stored and calibrated offline due to explicit cross correlations between all antenna pairs, which is not the case for gridded electric fields. Thus, direct imagers have the added requirement to calibrate in real-time since individual antenna information is not retained after gridding. \cite{beardsley_efficient_2017} has successfully demonstrated an algorithm for this purpose.

The E-Field Parallel Imaging Correlator (EPIC), a generic implementation and simulation of this imaging approach in Python, was described by Thyagarajan \citep{thyagarajan_generic_2017}. As a streaming, direct imaging correlator, it can be thought of as a generic, flexible real-time camera of the radio sky for large interferometer arrays.

Here, we report a GPU-accelerated implementation of EPIC, built on Bifrost, a high performance streaming framework.  The implementation has been deployed on the LWA station located on the Sevilleta National Wildlife Refuge in New Mexico, USA.  First light observations are shown, demonstrating its capability for transient detection. 

The theory of the MOFF (Modular Optimal Frequency Fourier) formalism underlying the EPIC imager is reviewed in Section \ref{sec:theory}, and a technical description of the implementation and development is discussed in Section \ref{sec:implementation}. First light observations and an initial
meteor transient detection are shown in Section \ref{sec:firstlight}, with benchmarks characterizing the performance on the LWA Sevilleta site discussed in Section \ref{sec:benchmarks}.  We summarize future work and conclude in Section~\ref{sec:conclusion}.

\section{Theory} \label{sec:theory}

The interferometry formulation is based on the optical theory of partially coherent quasi-monochromatic light, by the van Cittert-Zernike theorem \citep{zer38,born_principles_1999}. From this a relationship can be derived between the radiation pattern on the celestial sphere (in the far field) and a spatial coherence function measured on some plane between two points sampling the radiation pattern from the celestial sphere. This coherence function is the cornerstone of radio interferometry and is known as a `visibility'. 

A modern derivation can be found in \cite{thompson_interferometry_2017}, where the Fourier relationship between the sky co-ordinates and the interferometer co-ordinate system is described:

\begin{multline} \label{eq:vc}
 I(l,m,w) = \iint V(u,v,w) \\ \exp\bigg[2\pi i \big(ul + vm + w\big(\sqrt{1-l^2-m^2}-1\big)\big)\bigg]du\: dv \mbox{,}
\end{multline}

Mathematically this can be described by an outer-product between a vector representing a single frequency channel of fourier-transformed voltages from all antennas, and its conjugate transpose. Thus given $N$ antennas outputting $N$ electric field patterns in a channel, we derive a resultant $N \times N$ visibility matrix. Because of Hermitian symmetry, only the upper or lower triangle is retained for efficiency in FX correlators. This relation is as follows: 

\begin{align} \label{eq:vis}
  V_{12}(u,v,w) = E_{1}(x_1,y_1,z_1) \otimes E_{2}(x_2,y_2,z_2)^{\ast} \mbox{.}
\end{align}

Here $E(x,y,z)$ represents the electric field measured by an antenna at some location in an orthonormal co-ordinate system, with $V(u,v,w)$ representing the resultant visibility matrix. The (u,v,w) co-ordinate system represents the vector separation (baseline vector) between the different antennas.

\subsection{MOFF}\label{sec:MOFF}

The multiplication-convolution theorem from Fourier transform theory allows us to re-arrange Equation \ref{eq:vc}
to form the MOFF algorithm \citep{morales_enabling_2011} of

\begin{multline} \label{eq:vc2}
 I(l,m,w) = \Bigg\langle \bigg|\iint E(x,y,z) \times \\ \exp\big[2\pi i \big(xl + ym + z\big(\sqrt{1-l^2-m^2}-1\big)\big)\big]dx\: dy \bigg|^2 \Bigg \rangle \mbox{.}
\end{multline}

It is important to note that $E(x,y,z)$ constitutes the electric 
field in the Fourier domain convolved with the antenna illumination pattern. It is not a point function, but an electric field distributed across some 
physical extent in the measurement plane. Taking this into account Equation \ref{eq:vc2} becomes:

\begin{multline} \label{eq:vc_grid}
 I(l,m,w) = \Bigg\langle \bigg|\iint \big[W(x,y,z) \ast E^{\prime}(x,y,z)\big] \times \\ \exp\big[2\pi i \big(xl + ym + w\big(\sqrt{1-l^2-m^2}-1\big)\big)\big]dx\: dy \bigg|^2 \Bigg \rangle \mbox{,}
\end{multline}

where $W(x,y,z)$ defines a `gridding' function which constitutes
a convolution in antenna space, $E^{\prime}$ represents the point measurement of the electric field within the measurement plane, and $*$ the convolution operator. In addition to the antenna illumination pattern, $W(x,y,z)$ can optionally incorporate any wide-field effects resulting
from non-coplanarities in the array, as well as ionospheric effects \citep{morales_enabling_2011}. In our implementation, we assume a co-planar
array. Correcting for non-coplanarities will be dealt with in a forthcoming paper. With this in mind, Equation \ref{eq:vc_grid} becomes:

\begin{multline} \label{eq:vc_grid_coplanar}
 I(l,m) = \Bigg\langle \bigg|\iint \big[W(x,y) \ast E^{\prime}(x,y)\big] \exp\big[2\pi i \big(xl + ym\big)\big]dx\: dy \bigg|^2 \Bigg \rangle \mbox{,}
\end{multline}

Thus Equation \ref{eq:vc_grid_coplanar} is a gridding of an electric field pattern directly for each antenna, followed by a spatial Fourier transform to produce the image.  This transform is followed by squaring the image, or cross multiplying between polarisations, and accumulating images over time.  This produces what are commonly called `dirty images', which is the true sky brightness distribution convolved with the point spread
function of the instrument\citep{SIRA-II}.

\begin{figure}
  \centering
   \hspace*{-0.75cm}\includegraphics[trim=0cm 0 0 0,width=0.58\textwidth]{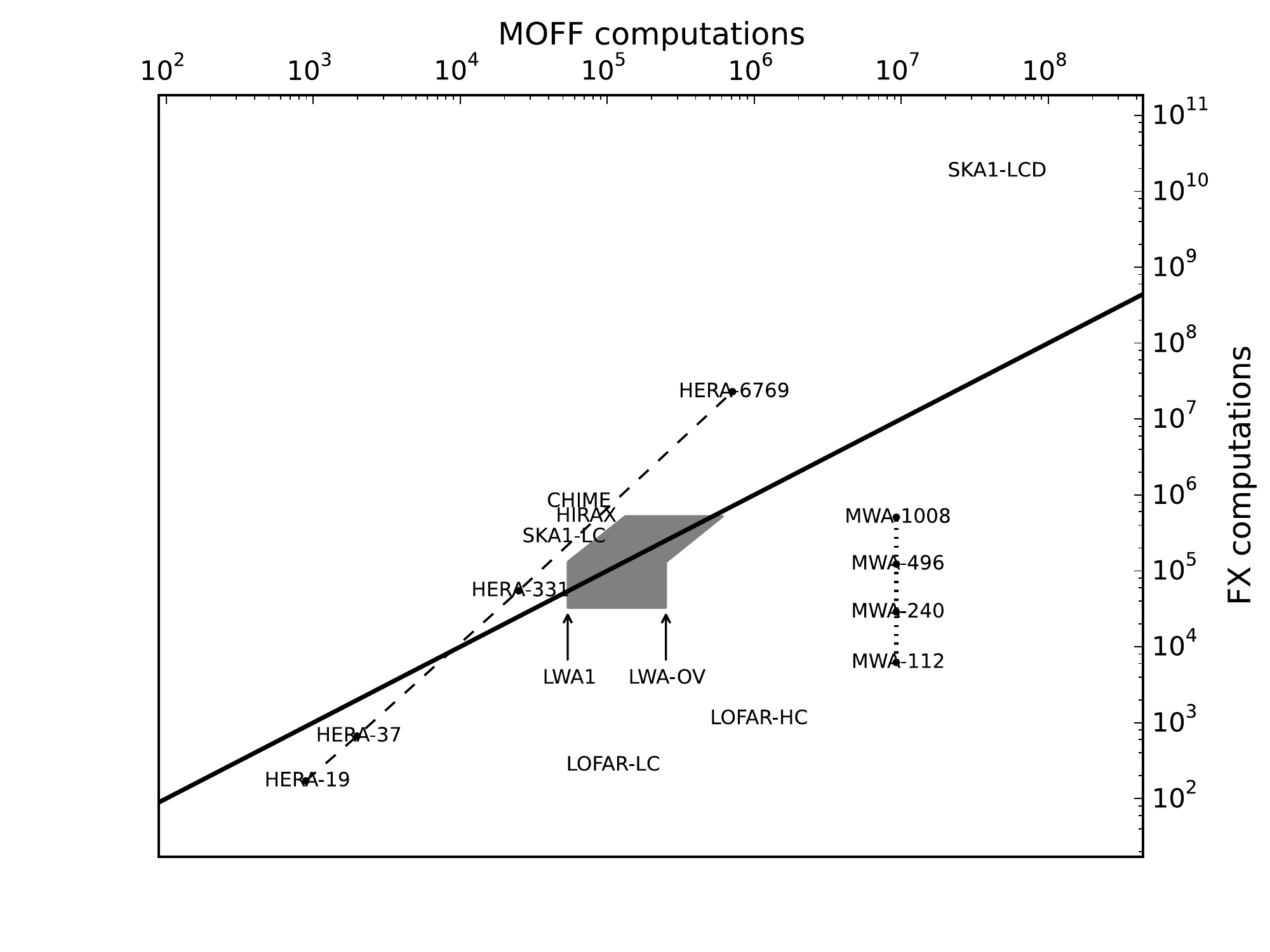}
  \caption{(Reproduced from \citet{thyagarajan_generic_2017})  A selection of current and planned instruments, with the solid black line delineating the boundary in efficiency between EPIC and FX correlators. Instruments below this line are more efficient with a standard FX correlator. Above the line it is more efficient to use EPIC.}
  \label{fig:moffvsepic}
\end{figure}

The EPIC architecture uses the MOFF algorithm as the basis for imaging. The computational cost of the EPIC architecture scales theoretically as $\mathcal{O}(n_g\log{n_g})$, compared to $\mathcal{O}(n_a^2)$ for the classical FX correlator. For highly dense arrays, depending on array geometry, a MOFF-based correlator, such as EPIC, may be more efficient than an FX correlator \citep{thyagarajan_generic_2017}. The limiting factor for the EPIC architecture is the Fourier transform size of the grid, whereas that for an FX correlator is the number of antenna pairs. A comparison of instruments and their most suited correlator type is shown in Figure \ref{fig:moffvsepic} \citep[reproduced from][]{thyagarajan_generic_2017}.

An additional characteristic of the EPIC architecture is the typically lower data rates in saving images at high time cadence \citep{thyagarajan_generic_2017}. Images from the EPIC architecture are already calibrated and science-ready compared to visibilities from an FX architecture which typically require additional processing offline to form science-ready images. This means that the sky can be imaged at a higher time resolution than is
possible using an FX correlator. The ramifications for EPIC, as will be seen, is that all-sky imaging at sub-millisecond sampling periods is feasible, potentially yielding new insights into a wide range of time-domain phenomena at radio frequencies.

\section{Implementation} \label{sec:implementation}

\subsection{Bifrost}

The GPU-accelerated implementation of the EPIC architecture is done using the Bifrost framework \citep{cranmer_bifrost:_2017}. Bifrost is a highly abstracted library for building high performance streaming systems. The back-end framework is built using C++ which calls high speed CUDA libraries
and bespoke kernels implemented by the Bifrost authors. For ease of use an abstracted Python front-end is provided.

Bifrost is based around the concept of blocks, where each block performs
some operation on the data, and then outputs it into a high-speed
ring buffer in memory, which facilitates moving data between blocks.
The output ring buffer from one block becomes the input ring buffer
for the next block. Each block loads a `gulp' of data from the ring, and processes it before placing a gulp into the output ring. The block processes data until the input ring is emptied or the pipeline is shutdown.

Many standard signal processing techniques are implemented into
the bifrost back-end with GPU capability where appropriate. These
include operations such as finite impulse response filters, fast Fourier transforms (FFTs), and various matrix algebra operations.

The Python front end also includes a high performance mapping
function which takes a string of C++/CUDA and uses a Just-In-Time (JIT) compiler to generate and execute valid CUDA code on-the-fly using
the Bifrost back-end. This provides significant flexibility in 
doing small mathematical operations without the need to write
multiple custom blocks and implement them directly into the Bifrost framework.

The majority of the EPIC implementation in Bifrost was done using the
standard signal processing blocks as well as the Bifrost map function,
with a notable exception of high-speed convolution and gridding.  For this operation a custom kernel was created based on a high speed gridder
developed by \cite{romein_efficient_2012}. 

\subsection{Pipeline}\label{sec:EPIC-arch}

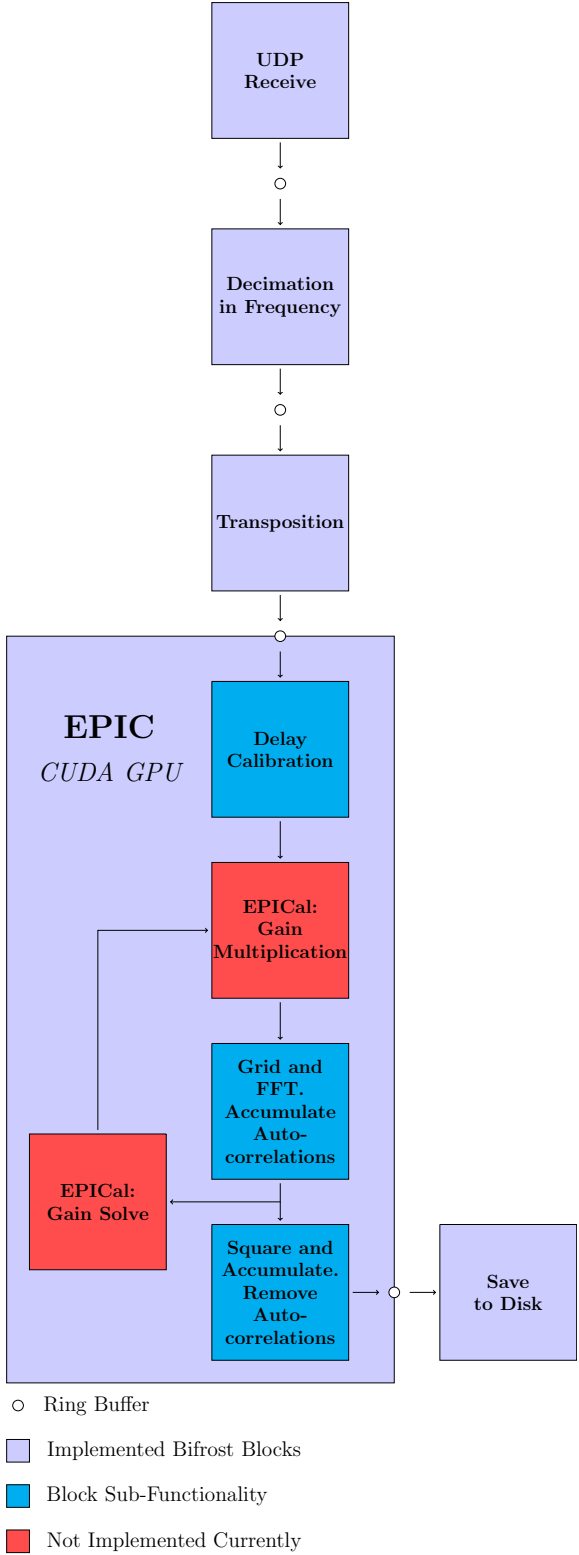
\begin{figure}
  \centering
  \scalebox{0.5}{\begin{tikzpicture}[x=1.2cm,y=1.2cm]

  \tikzstyle{every node}=[font=\LARGE \bfseries]

    \draw [fill=blue, fill opacity = 0.2] (0.0,15.0) -- (3.0,15.0) -- (3.0,15.0 - 3) -- (0.0,15.0 - 3) -- cycle;
    \node at (1.5,13.5) [align=center] {UDP \\  Receive};

   \draw [fill=blue, fill opacity = 0.2] (0.0,10.0) -- (3.0,10.0) -- (3.0,10.0 - 3) -- (0.0,10.0 - 3) -- cycle;
  \node at (1.5,8.5) [align=center] { Decimation \\ in Frequency};

  \draw [fill=blue, fill opacity = 0.2] (0,5.0) -- (3.0,5.0) -- (3.0,5.0 - 3) -- (0,5.0 - 3) -- cycle;
  \node at (1.5,3.5) [align=center] { Transposition};
  
  \draw [fill=blue, fill opacity = 0.2] (4.0,1.0) -- (4.0,-15.5) -- (-4.5,-15.5) -- (-4.5,1.0) -- cycle;

  \draw [fill=cyan, fill opacity = 1.0] (0,0) -- (3.0,0) -- (3.0,0- 3) -- (0,0 - 3) -- (0,0);
  \draw [fill=white!30!red, fill opacity = 1.0](0,-4.0) -- (3.0,-4.0) -- (3.0,-4.0- 3) -- (0,-4.0 - 3) -- cycle;
  \foreach \i in {-2,...,-3}
  {
    \draw [fill=cyan, fill opacity = 1.0] (0,\i*4) -- (3.0,\i*4) -- (3.0,\i*4 - 3) -- (0,\i*4 - 3) -- (0,\i*4);
  }
  
  \foreach \i in {3,...,1}
  {
    \path (1.5,\i*5 - 3) node (y1) {} (1.5,\i*5-3.75) node (y2) {};
    \path (1.5,\i*5 - 4.25) node (y3) {} (1.5,\i*5-5) node (y4) {};
    \draw (1.5,\i*5 - 4) node[circle,style={minimum width=.5ex,fill=white},draw]{};
    \draw [->] (y1) -- (y2);
    \draw [->] (y3) -- (y4);
  }
  
  \foreach \i in {0,...,-2}
  {
    \path (1.5,\i*4 - 3) node (y1) {} (1.5,\i*4-4) node (y2) {};
    \draw [->] (y1) -- (y2);
  }

  \node at (1.5,-1.5) [align=center] { Delay \\ Calibration};
  \node at (1.5,-5.5) [align=center] { EPICal: \\ Gain \\Multiplication};
  \node at (1.5,-9.5)  [align=center] { Grid and \\ FFT. \\ Accumulate \\ Auto-\\correlations};
  \node at (1.5,-13.5)  [align=center] { Square and \\ Accumulate. \\ Remove \\ Auto- \\correlations};

  \draw [fill=blue, fill opacity = 0.2] (5.0,-12) -- (8.0,-12) -- (8.0,-15) -- (5.0,-15) -- (5.0,-12);
  \node at (6.5, -13.5) [align = center] { Save \\ to Disk };

  \draw [fill=white!30!red, fill opacity = 1.0] (-1.0,-10) -- (-4.0,-10) -- (-4.0,-13) -- (-1.0,-13) -- (-1.0,-10);
  \node at (-2.5, -11.5) [align = center] { EPICal: \\ Gain Solve };
  \path (1.6,-11.5) node (e1) {} (-1.0,-11.5) node (e2) {};
  \draw [->] (e1) -- (e2);

  \path (3.0,-13.5) node (s1) {} (3.75, -13.5) node (s2) {};
  \path (4.25,-13.5) node (s3) {} (5.0, -13.5) node (s4) {};
  \draw (4.0, -13.5) node[circle,style={minimum width=.5ex,fill=white},draw]{};
  \draw [->] (s1) -- (s2);
  \draw [->] (s3) -- (s4);

  \path (-2.5, -10) node (epical1) {} (0,-5.5) node (epical2) {};
  \draw [->] (epical1) |- (epical2);
 \node at (-2.25,-1) [align= center, font=\fontsize{25}{1}\selectfont \bfseries] {EPIC};
  \node at (-2.25,-2) [align= center, font=\fontsize{20}{1}\selectfont, thick] {\it CUDA GPU};


  \draw (-4.25, -16) node[circle,style={minimum width=.5ex,fill=white},draw]{};
  \draw (-2.45, -16) node [font=\fontsize{16}{1}\selectfont,align=left]{ Ring Buffer };
  \draw [fill = blue, fill opacity = 0.2] (-4.5, -16.75) -- (-4.0,-16.75) -- (-4.0,-17.25) -- (-4.5,-17.25) -- cycle;
  \draw [fill = cyan, fill opacity = 1.0] (-4.5, -17.75) -- (-4.0,-17.75) -- (-4.0,-18.25) -- (-4.5,-18.25) -- cycle;
  \draw [fill = white!30!red, fill opacity = 1.0] (-4.5, -18.75) -- (-4.0,-18.75) -- (-4.0,-19.25) -- (-4.5,-19.25) -- cycle;
  \draw (-0.75, -17) node [font=\fontsize{16}{1}\selectfont,align=left]{ Implemented Bifrost Blocks };
  \draw (-1.2, -18) node [font=\fontsize{16}{1}\selectfont,align=left]{
  Block Sub-Functionality};
  \draw (-0.75, -19) node [font=\fontsize{16}{1}\selectfont,align=left]{ Not Implemented Currently };

\end{tikzpicture}}
  \caption{Block diagram of the Bifrost-based implementation of the EPIC architecture at LWA-SV.  The blocks are named by their function and the arrows indicate the direction of data flow. The large EPIC block corresponds to a single operational block in the bifrost pipeline, with its major sub-functionality displayed. Where the calibration steps sit are also included, despite not yet being implemented. The EPIC block maps closely with the architecture discussed in \citet{thyagarajan_generic_2017} .}
  \label{fig:pipeline}
\end{figure}

The real-time streaming processor implementation comprises several Bifrost blocks\footnote{The source code for EPIC as well as the Bifrost-based pipeline implementation for the LWA is available at \url{https://github.com/nithyanandan/EPIC}.} as a Python program, with all significant compute and memory operations done seamlessly through Bifrost's high performance C++/CUDA backend. An overview of this pipeline and the relationship between the various blocks is shown in Figure \ref{fig:pipeline}.

Channelized raw data is received via UDP in a 4+4-bit complex integer format in the {\tt UDP Receive} block.  This complex integer representation serves to reduce the bandwidth required by the local ethernet connection.  After the data has been captured, the channelised data is first decimated in frequency to obtain a bandwidth that can be processed without packet loss, and then transposed to move the data ordering from $\big[Time, Channel, Antenna, Polarisation\big]$, to $\big[Time, Channel, Polarisation, Antenna\big]$. This is important as it facilitates contiguous loads in the gridding convolution step, discussed below.

After the {\tt Transposition} block, the complex integer data is
unpacked and promoted to a standard 64-bit complex floating-point number (32-bit real, 32-bit complex) and compensation for the signal path delays are applied using a JIT compiled Bifrost {\tt map} function. The delay calibrated data is then convolved with the antenna illumination pattern, which is a user-defined convolution kernel. This is then gridded onto a 2D grid with a spacing of $<\frac{\lambda}{2}$, where $\lambda$ is the wavelength of the channel, to ensure we are sampling all of the sky modes by sampling at the Nyquist wavelength.

This convolution and gridding operation is done using the Romein Convolution algorithm \citep{romein_efficient_2012} in the {\tt Grid and FFT} block in Figure \ref{fig:pipeline}, designed specifically for high speed visibility gridding where locality is poor and memory bandwidth is high. The gridding convolution algorithm is described in more detail below.

Once the data has been gridded for a single time step, the gridded data is inverse Fourier transformed to produce a complex-valued image on the sky. These images are then cross-multiplied in the {\tt Square and Accumulate Image} block to form the polarised images, which are then accumulated to a user-defined time interval depending on the science use case.  After accumulation to the threshold time, the image is written to disk in a binary format and converted to a FITS image in a post-processing step. This ensures the real-time processing is not held up by high-cost image manipulation operations.

Optionally autocorrelation removal can also be done to remove the zero-spacing power inherent in EPIC. Together with this, the imprint of the image of the gridding illumination kernels can be removed after the fact in a post-processing step as they are pre-generated and thus known previously.

\subsection{Romein Convolution Algorithm}

The Romein convolutional algorithm \citep{romein_efficient_2012} proved to be a critical step in the implementation of EPIC. Previous EPIC reference codes have attempted to use a direct convolution mapping using matrix multiplication, as described by the operator formalism in \cite{thyagarajan_generic_2017}.

Unfortunately, on a GPU this results in unacceptably high memory bandwidth which causes this step to bottleneck the code significantly. The Romein convolution was used instead as it is designed to reduce the GPU memory bandwidth significantly by only doing explicit memory store operations when necessary. The algorithm is designed to preferentially accumulate any grid
updates into a high speed local register on the GPU core.

The Romein convolution algorithm additionally allows multiple convolution kernels to be
combined together and applied simultaneously. This not
only allows convolution of the electric field with the
illumination pattern, i.e., A-projection \citep{morales_software_2009,bhatnagar_correcting_2008}, but additionally provides scope for including wide-field
and antenna effects, such as W-Projection \citep{cornwell_noncoplanar_2008}. 
The implementation of non-coplanarity correction to ensure
wide field fidelity will be a focus of future work.

The Romein convolution  algorithm, written in C++/CUDA, was implemented by modifying the Bifrost
back-end and to add the necessary functionality. The additional Bifrost module is intended to be a generic, type-agnostic convolution kernel. This module is then called in the pipeline script from the Bifrost library using Python's {\tt ctypes} interface.

\section{Deployment and First Light} \label{sec:firstlight}

\subsection{Long Wavelength Array}

\begin{figure}
  \centering
  \includegraphics[width=0.5\textwidth]{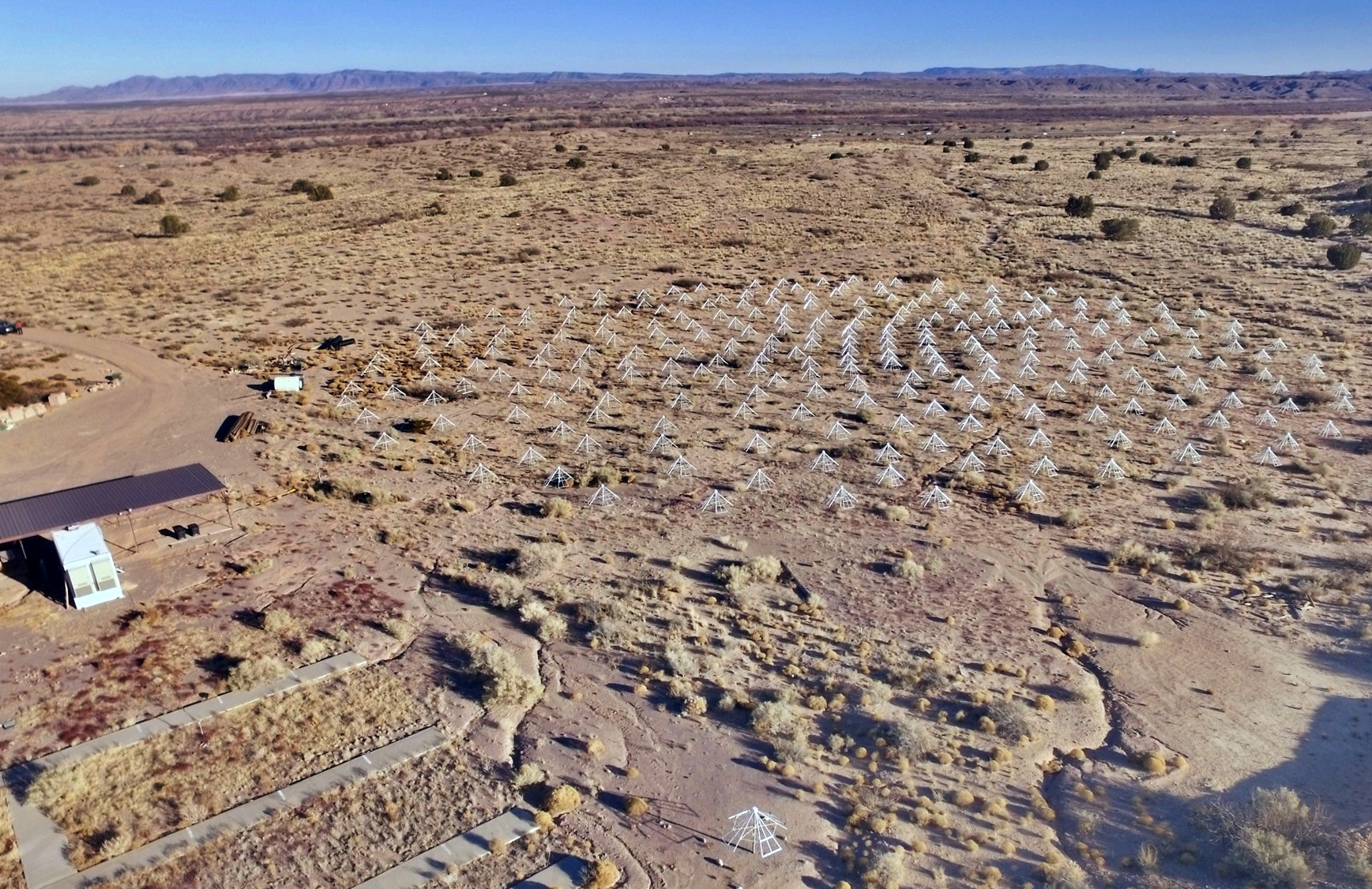}
  \caption{Aerial view of the LWA station at the Sevilleta National Wildlife Refuge. Most antenna elements are in a dense configuration towards the right of the image. A test antenna, is visible at the bottom. The signal processing hardware is contained within a modified, radio frequency shielded shipping container, visible in the left of the image.}
  \label{fig:sev}
\end{figure}

The LWA is a low frequency radio interferometer observing between frequencies of 10 to 88 MHz, with two operational stations, one located at the Karl G. Jansky Very Large Array site and the other at the Seviletta National Wildlife Refuge (see Figure~\ref{fig:sev}), both in the state of New Mexico, USA \citep{henning_first_2010,taylor_first_2012,ellingson_taylor_2013}. Its high density configuration makes it an excellent candidate for deployment of EPIC.

The LWA Sevilleta (LWA-SV) array consists of 256 dipole antennas arranged in a dense pseudo-random arrangement inside a 110~m by 100~m elliptical aperture that is elongated north-south. An additional antenna is located approximately 300 m west of the core of the array, acting as an outrigger to help with calibration and to improve the angular resolution of the telescope. This outrigger was explicitly excluded during our implementation to ensure a high density, keeping the resultant image FFT size as small as possible.

The analog signal from each dipole is initially low pass filtered and amplified at the front end before being transmitted over coaxial cable to the electronics shelter.  Inside the shelter the analog signal is further filtered and then digitized using ROACH2 boards.  The boards use the CASPER ADC16x156-8 digitiser boards to sample the dipole signals at 204.8 MHz.  The digitized signals are then Fourier transformed into 4096 25 kHz channels with a time resolution of 40$\mu$s. At this point the frequency domain data, between 10MHz and 88 MHz, are requantised into 4+4-bit complex integer data, packetised, and routed over a 10/40 GbE network to a cluster of seven general purpose machines.  Each machine is equipped with two Intel Xeon E5-2640 v3 processors, 64 GB of RAM, a Mellanox ConnectX-3 40 GbE network interface card, and two NVIDIA GTX 980 (Maxwell) GPUs.

\subsection{Deployment}

The initial deployment took place on the LWA-SV site during the week of the 27-31 August 2018. The EPIC architecture was deployed on a single cluster node, receiving a sixth of LWA-SV's total bandwidth.  Operation of EPIC was achieved with no modifications to the LWA system or hardware apart from swapping the FX correlator software pipeline for EPIC.  The LWA's public software library was used to perform delay calibration to account for different antenna cable lengths and to provide the array geometry \citep{dowell_long_2012}.  A simple square top-hat function with 3-meter extent was used as the illumination pattern for the dipole antennas.  No additional calibration was performed.  The observations reported here were run at an image accumulation time of 50~ms in order to allow observations of short-duration transient phenomena in the radio sky.  Four channels of 25~kHz were processed with a combined bandwidth of 100~kHz centred at a frequency of 55.25~MHz.  The stability of the system was tested with a 24 hour operation under the EPIC correlator.  Images were generated at the raw 40~$\mu$s time cadence of the LWA-SV and then accumulated to obtain the final cadence of 50~ms.  A $\lambda/2$ grid spacing was used, resulting in approximately $64^2$ image pixels.

\subsection{Detection of Meteor Transient as Proof of Concept}

EPIC images the whole sky as visible to the LWA-SV station. During our initial observations, multiple small transients were identified. The majority of them are radio frequency interference (RFI), which most often shows up on the horizon, indicating a terrestrial origin. Occasionally RFI can appear overhead, reflected off of airplanes or satellites. These signals are generally narrow bandwidth and highly polarized, making them easy to recognize.

\begin{figure}
  \centering
  \includegraphics[width=0.4\textwidth]{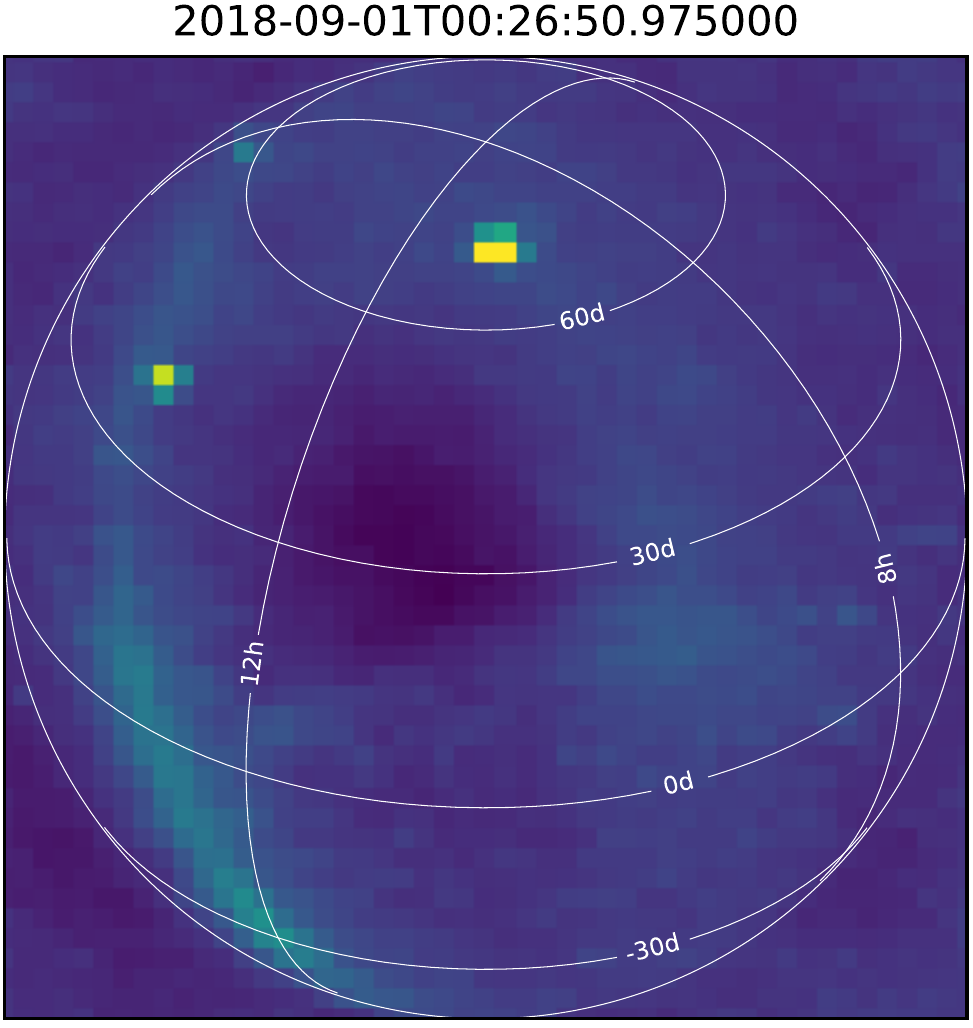}
  \caption{All-sky pseudo Stokes-I image showing a meteor reflection detection during an observation on the LWA-SV site (upper center). The plasma left by the meteor impacting the atmosphere reflects the signal from a 55.25 MHz TV transmitter located beyond the horizon.  Lines of constant right ascension and declination in J2000 are marked in white.  Cygnus A is the bright point in the upper left of the image. A .mp4 video file of this event is available, and has been submitted
  alongside this manuscript. The video is of images outputted sequentially at a 50ms cadence.}
  \label{fig:meteor}
\end{figure}

After ruling out RFI events, some physical transients were noted, the
brightest of which in our observing window was a meteor striking
the Earth's atmosphere, a still frame pseudo Stokes-I image of which is shown in Figure~\ref{fig:meteor}. A pseudo-stokes image is one that is formed from straightforward linear combinations of the coherency vectors from the linear polarisation parameters, but is acknowledged to not exactly represent the true stokes vectors due to cross-coupling and polarisation leakage effects. The meteor striking the atmosphere generates a plasma, which acts as a reflector for an over the horizon analog TV transmitter at 55.25~MHz, illuminating the meteor plasma's path. This is almost identical to the methodology of studying meteor events through the use of radar \citep{1947MNRAS.107..155P}.  Studies of reflections such as these provide information about the speed of the neutral wind in the mesosphere and lower thermosphere through the observed Doppler shift of the reflection \citep{helmboldt_ellingson_2014}.  The total number of meteor reflections can also be used to inform estimates of the terrestrial accretion rate \cite[see][and references therein]{accretion}.  Such events have been observed by the LWA previously, as well as self-emission from meteor trails \citep{obenberger_detection_2014} and lightning \citep{obenberger_using_2018}. This demonstrates the potential of an EPIC system for image-based all-sky transient detection and monitoring. 


\begin{figure}
  \centering
  \includegraphics[width=0.5\textwidth]{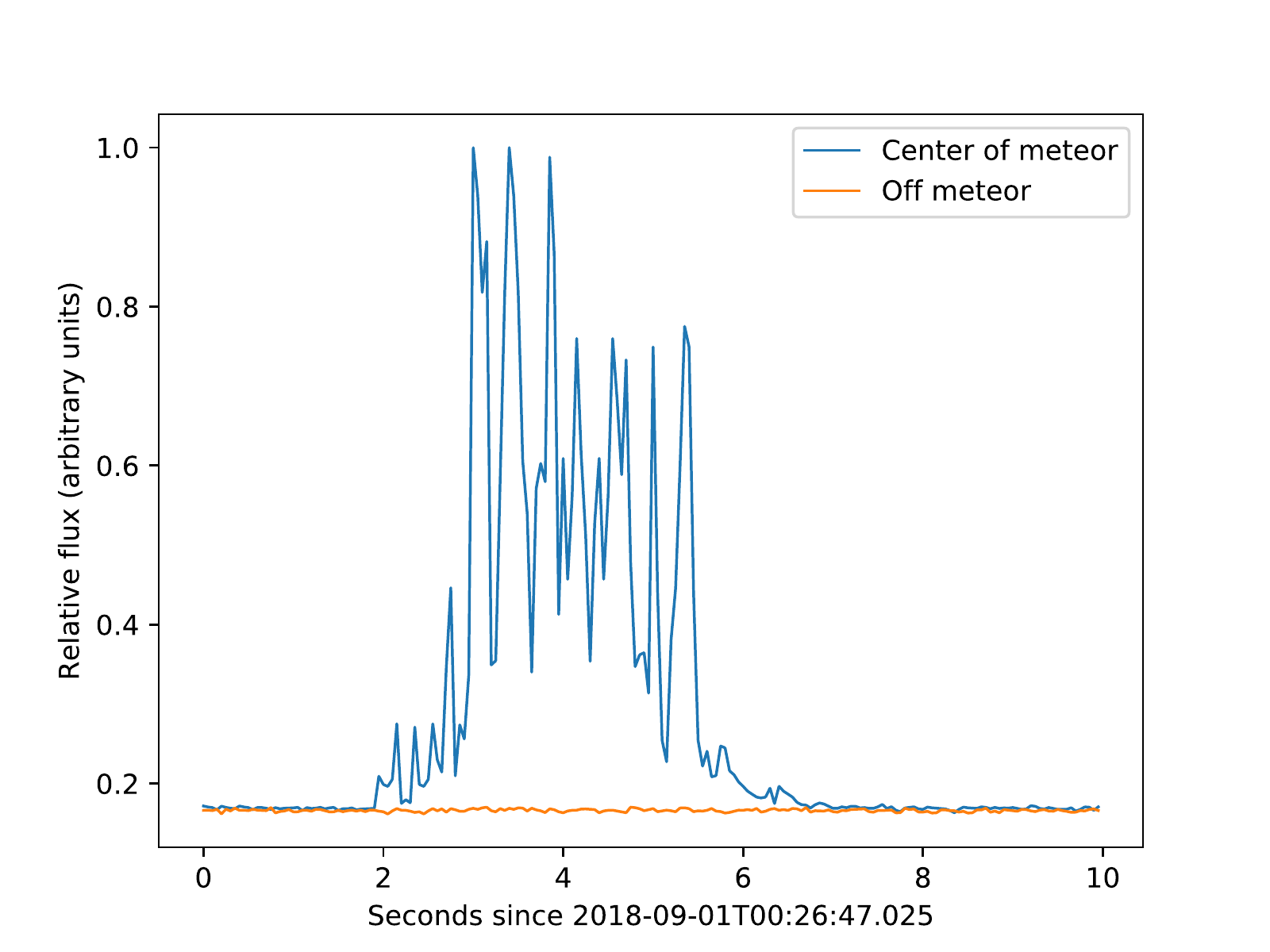}
  \caption{Light curve of the brightest pixel around the transiet during the meteor passage noted in Figure \ref{fig:meteor}, with a comparison to the radio background.  The time resolution is 50 ms.  The reflection lightcurve shows considerable structure due to changes in the plasma tail as it expands and is distorted by atmospheric winds. The lightcurve is consistent with some of the examples in \citet{helmboldt_ellingson_2014}.}
  \label{fig:meteor_lightcurve}
\end{figure}

\section{Benchmarks} \label{sec:benchmarks}

During the first light deployment at the LWA-SV site, the performance was measured and characterised. The performance is a consequence of both the deployment system and hardware, as well as EPIC's execution method in comparison to an FX correlator.

Overall, in the first iteration, up to 800 kHz of bandwidth is processed per GPU card on the LWA-SV correlator, when running with only a single instrumental polarisation, which is useful for maximising bandwidth for faint transients and facilitates averaging over the band. With the LWA-SV system's current hardware layout, this corresponds to 9.6 MHz of single polarisation bandwidth when EPIC is run on both GPUs of all six data capture servers.  When running with both X and Y polarisations, which allows the formation of Stokes images, half the overall bandwidth is available: up to 400 kHz per card or 4.8 MHz for the entire system. We explore the factors contributing the per-GPU bandwidth below and discuss ideas for improvement.

\subsection{Maximum Throughput}

To characterise the overall throughput of the system, we monitored the UDP streams being broadcasted by the ROACH2 boards running the front-end Fourier transforms and channelisation. If the system is keeping up with the input data, then there will be no packet loss. If compute requirements increase on the node, for example by increasing the number of channels per card or changing the frequency tuning such that a larger grid/FFT size is needed, then packet loss will occur as the pipeline struggles to keep up with the incoming data stream.  

There are additional overheads in the system, such as running a normal Linux operating system in the background that can cause the occasional reductions in processing performance.  To ensure that the pipeline does not drop packets due to such variations, we found empirically that a time `gulp', i.e, the amount of time represented by a single chunk of data, such that the data can be processed in $\approx$90\% of the observed time is useful, providing a 10\% margin for system processing variations. For example, if ingesting 50~ms worth of data in a single gulp from the ring buffer, to ensure the system can keep up, the GPU should process it in 45~ms to keep the system running smoothly.

\begin{figure}
  \centering
  \includegraphics[width=\columnwidth]{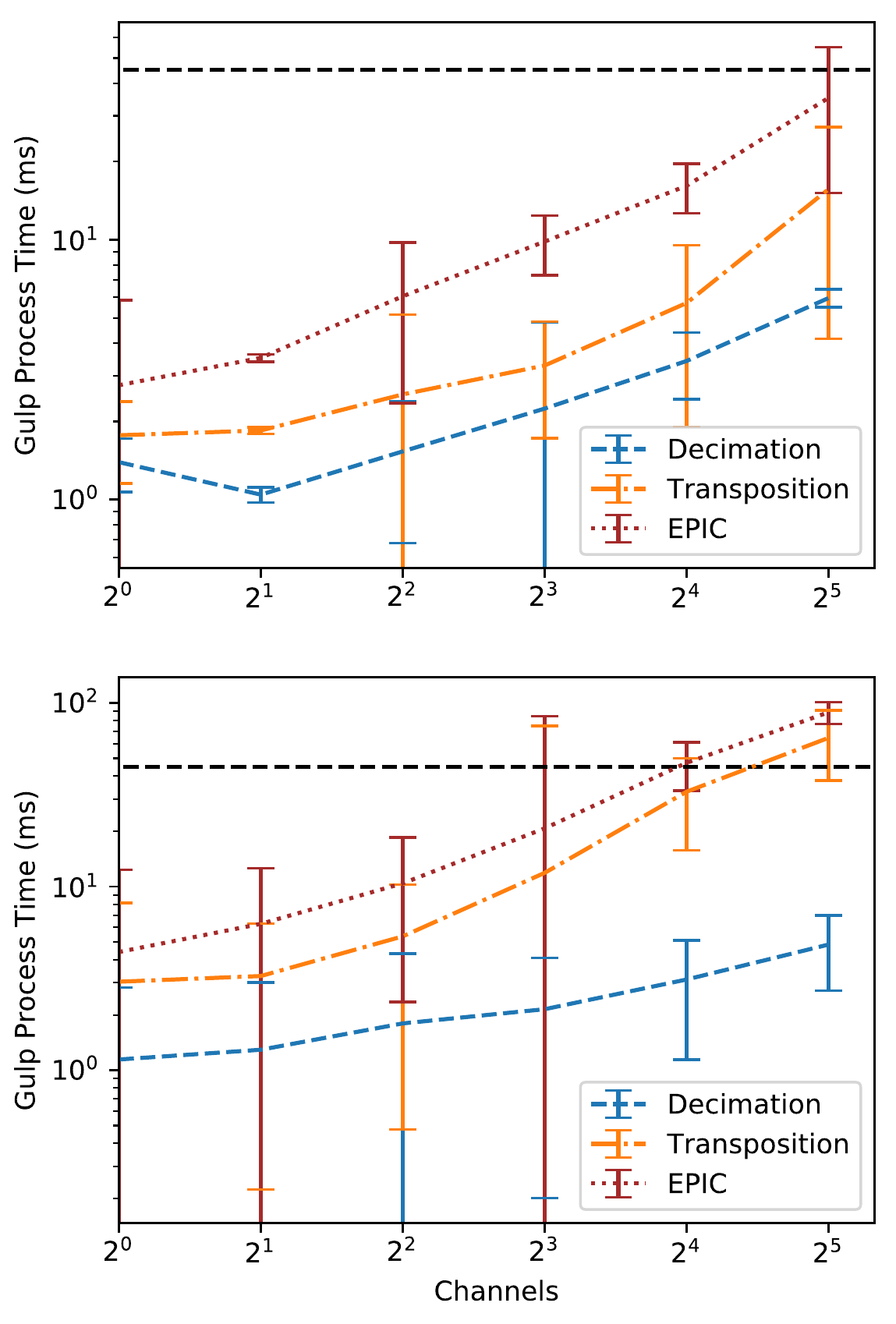}
  \caption{Processing time as a function of the number of channels being processed for (top) single polarisation and (bottom) dual polarisations, with EPIC running on LWA-SV with a time gulp of 50 ms and grid size of 32 pixels on a side. At 32($2^{5}$
  channels is when we began experiencing packet loss on dual polarisations, on the incoming data stream carrying electric field data, which marks when the system is no longer able to keep up with the input data rate. The black dashed horizontal line denotes the 90\% processing time for the gulp size.}
  \label{fig:throughput}    
\end{figure}

The results of our initial tests on the system are shown in Figure~\ref{fig:throughput} where the gulp processing time and UDP packet loss fraction are shown as a function of the number of frequency channels processed.  As computational resources are exceeded by increasing the number of channels, the pipeline backs up, and packet loss increases to indicate that system capacity has been exceeded.

\begin{figure}
    \centering
    \includegraphics[width=\columnwidth]{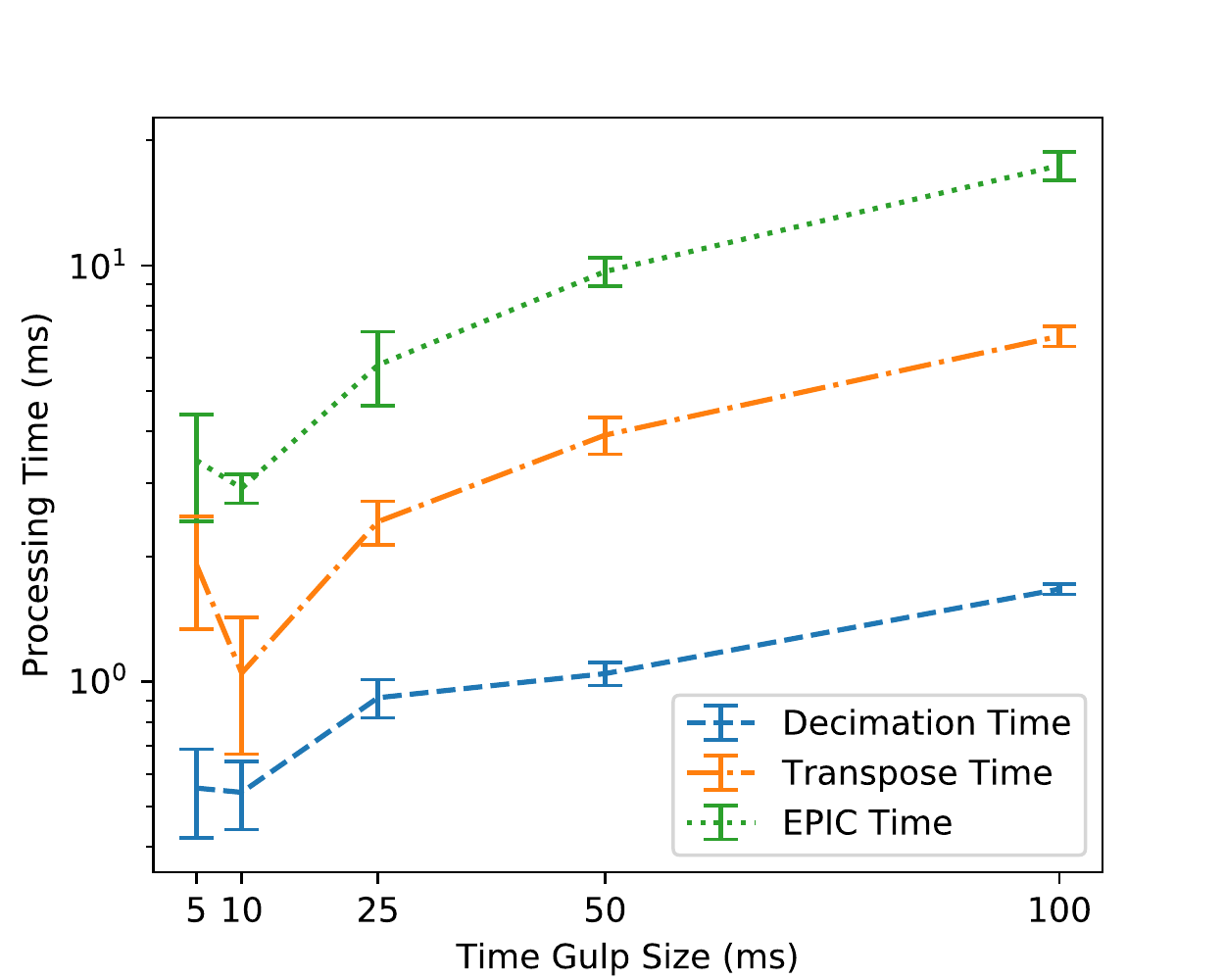}
    \caption{An exploration of how the system scales as a function of the time gulp sizes for 100 kHz of bandwidth and dual polarization.  Each vertical bar is sub-divided to show the time used by each block in the pipeline. The legend corresponds to the blocks in Figure \ref{fig:pipeline}, with `Image and Accumulation' corresponding to the pipeline element on the CUDA GPU. This data was derived from at least 600 trials of each time gulp size. }
    \label{fig:time_scaling}
\end{figure}

\begin{figure}
    \centering
    \includegraphics[width=0.5\textwidth]{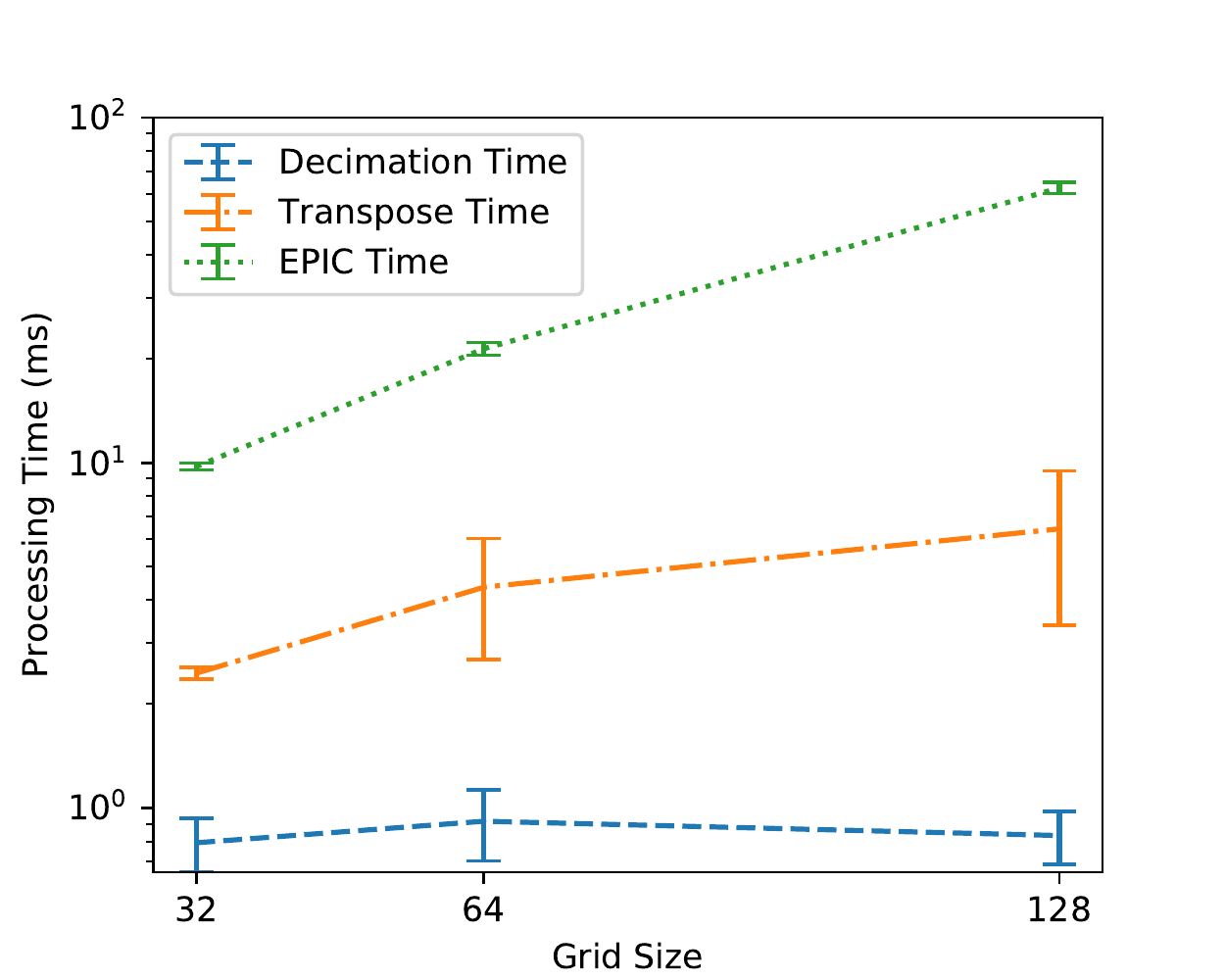}
    \caption{An exploration of how the system scales as a function of the grid size for a time gulp of 25 ms, 100 kHz of bandwidth, and dual polarization. The Grid Size is the size of one dimension of our squared grid. This data was derived from at least 600 trials of each grid size. 'EPIC Time' in this instance is the GPU element specified in Figure \ref{fig:pipeline}. A grid size of 128 is more than can be processed in real-time by the system, as it causes packet loss on the input UDP stream, but it is plotted here to show scaling.}
    \label{fig:grid_scaling}
\end{figure}

With a grid size of $64^2$ and the time gulp size set to 50ms, we are capable of running up to 16 channels (400 kHz) with dual polarisations before packet loss increases to indicate the pipeline stalling. Single polarisation mode runs over twice as fast as the dual polarisation mode.
 In Figure~\ref{fig:time_scaling}, the scaling of the system as a function of time gulp size is shown when processing 100 kHz of bandwidth and dual polarisation. The scaling with gulp size is roughly linear, with the GPU coping well at a variety of representative time gulp sizes between 5 ms and 0.1 s. Similarly, Figure~\ref{fig:grid_scaling} shows the scaling of the pipeline with the grid size for a gulp size of 25 ms with the same bandwidth and polarization setup as in Figure~\ref{fig:time_scaling}.  We see that the processing times for grid sizes of 32 and 64 pixels on a side are roughly comparable, indicating that the EPIC processing time may not be dominated by the Fourier transform at these grid sizes.  This can also been seen in Table~\ref{tab:block_ratios} where the processing time per gulp is explored for a representative pipeline run.  The scaling of the processing time between 64 and 128 pixels on side is around 2.5 times whereas theory would predict an increase of 4.7. The reason for this is potentially the underlying cuFFT library being more efficient for larger FFT sizes compared to smaller ones \citep{kent_2016}. We note that larger sizes for both the time gulp and grid size are unable to be tested because of the lack of sufficient memory on the GTX 980 GPUs avaliable at LWA-SV.

\subsection{GPU Performance}

Here we assess the overall performance and suitability of EPIC for a GPU programming model. This can be explored using a roofline model, a common visualisation in high performance computing to analyze the execution properties of a particular algorithm \citep{demmel_roofline:_2008}.

\begin{figure}
  \centering
  \includegraphics[width=0.5\textwidth]{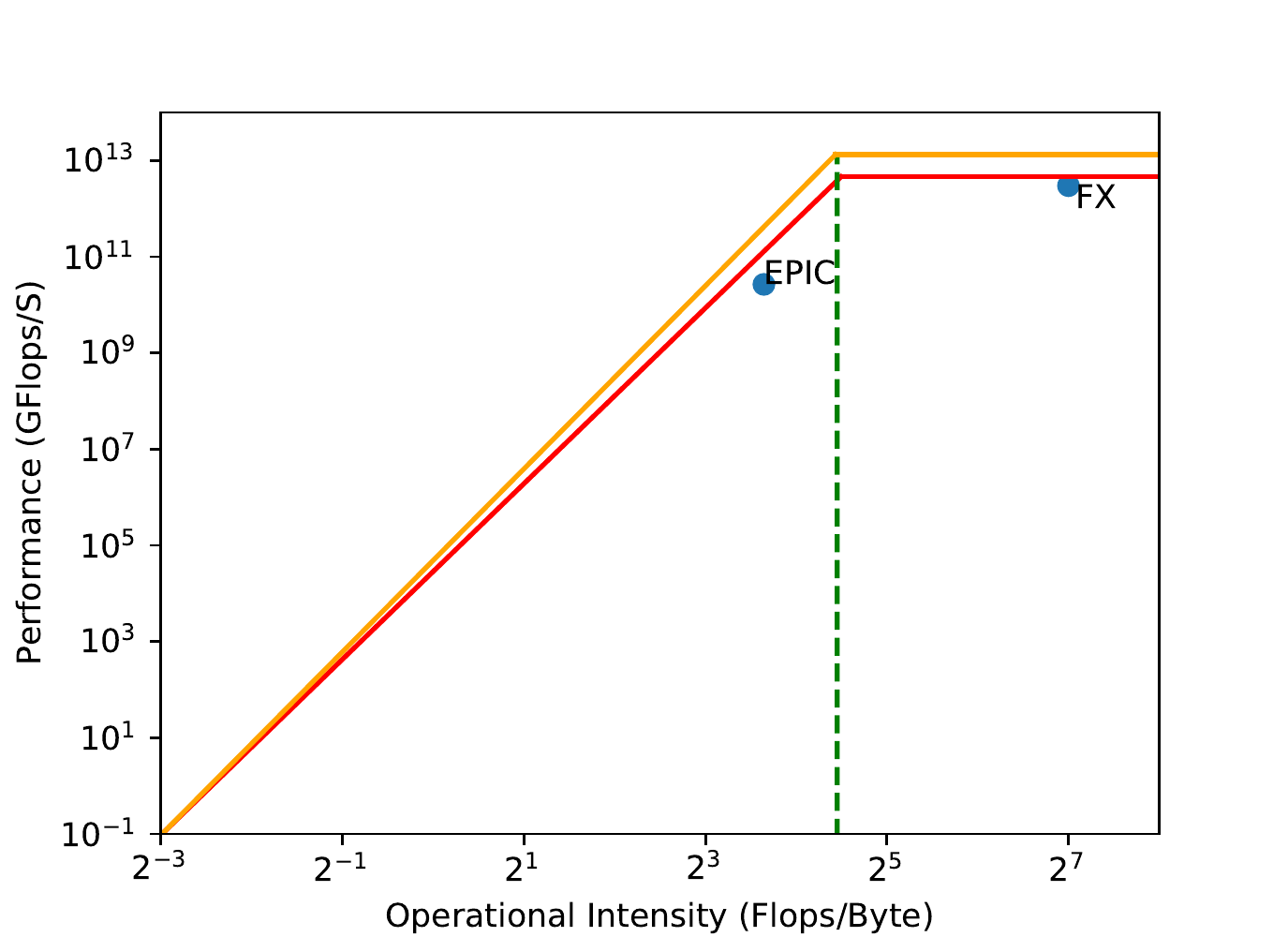}
  \caption{A roofline model comparing an FX correlator outer product with the EPIC pipeline on a NVIDIA GTX 980 GPU.  The red line represents the maximum number of operations per second for a GTX 980 GPU, and the orange line for a GTX 2080, a potential upgrade card for the LWA.  The rooflines for both cards were worked out using the memory bandwidth and peak compute performance for single-precision floating point (FP32) operations. The memory bandwidth of the GTX 980 and GTX 2080 are 226 GB/s and 616 GB/s, with peak FP32 performance of 4.6 and 13.3 TeraFlops, respectively.}
  \label{fig:epic_roofline}
\end{figure}

The roofline comparison between the GPU computed elements of an
FX correlator and EPIC is shown in Figure~\ref{fig:epic_roofline}. The 
example here is computed using a representative roofline for an NVIDIA GTX 980 GPU, used in this implementation, and for the elements of the pipeline that execute on the GPU. A GTX 2080 roofline is also provided, as a potential upgrade for the LWA correlator. The FX correlator is clearly in the compute bound regime, whereas EPIC is memory bound. This means increasing the memory bandwidth available rather than compute power will be more beneficial for EPIC, in contrast to a FX correlator which is predominantly compute-bound. 

Upgrading the LWA-Sevilleta correlator to use GTX 2080's, which have over double the memory bandwidth and compute performance, should yield a significant performance increase in bandwidth that can be processed. Assuming an increase in performance of at least two times with the new cards, not withstanding additional optimisation, over 15 MHz of LWA Bandwidth should be able to be processed.

We note two important caveats that such a direct comparison is not entirely appropriate. Firstly, the EPIC architecture provides end-to-end real-time imaging (from raw antenna voltages to science-ready calibrated images), whereas an FX correlator predominantly consists of a single mathematical operation, namely, outer product of the raw voltages and thus does not calibration or imaging, which incur additional costs. Secondly, if fast time-domain studies (timescales $\lesssim 1$~ms) are to be performed with an FX-based correlator, the cost of gridding and imaging will be much higher since they have to be performed at such fast cadences and have not been included in these estimates. 

\begin{table}
\centering
\begin{tabular}{ll}
Block & Processing Time \\
~ & (\% of gulp time)\\
\hline \hline 
Decimation & 2 \\
Transposition & 4 \\
EPIC & 90 \\
\hline
EPIC -- FFT & 35 \\
EPIC -- Gridding & 20 \\
EPIC -- Cross-Multiply & 40 \\
EPIC -- Data Transformation & 5 \\
\hline
Save & 4 
\end{tabular}
  \caption{Representative approximate breakdown of processing time by block as a fraction of the time gulp. This was done for a grid size of $64^2$, 2048 40$\mu$s time samples and eight channels.  }
\label{tab:block_ratios}
\end{table}

\section{Conclusion} \label{sec:conclusion}

The first version of a working EPIC direct imager, through direct implementation of the MOFF algorithm, has been developed fully, and its implementation and operation demonstrated at the LWA-SV site. Observations of transients from reflections of terrestrial transmissions off passing meteors on timescales of $\sim 2$~s at a cadence of 50~ms are reported. These serve to verify EPIC as a science-capable interferometric imaging capability.  

The Bifrost framework aided in implementing EPIC. The C++/CUDA back-end abstracts away complicated constructs, such as the ring buffers, which form the communication backbone between processing steps. The major advantage is the native CUDA support, facilitating access to the power of the GPGPU paradigm. Extending the Bifrost framework, such as adding extra GPU-enabled processing blocks, was straightforward.


This successful deployment and working demonstration of the principles behind EPIC and the MOFF formalism mark a paradigm shift in correlator technology. The impact is especially acute for high density arrays such as SKA1-Low and the completed HERA configuration. It can also offer the capability, with its next iteration of development as a self-triggering transient survey instrument for arrays such as the Low Band Observatory \citep[ngLOBO;][]{taylor_ngLOBO}, and the LWA Swarm Telescope  \citep{dowell_swarm_2018}. Higher frequency instruments such as the MWA can benefit from the unique capabilities of EPIC for exploring FRB phenomena, with its ability to image the entire celestial hemisphere simultaneously at high time cadence.  Additional potential scientific uses range from ionospheric disturbance mapping to direct observations of compact astrophysical sources. 

While the first version of the EPIC system is now operational, significant improvements are planned.  Future work will include:
\begin{itemize}
\item EPICal - EPIC requires calibration in real-time. A solution has been demonstrated \citep{beardsley_efficient_2017} but not implemented yet in this deployment. This will be an important feature addition since direct imaging approaches do not allow for post acquisition methods to improve the image calibration.
\item Wide Field Effects - Effects of non-coplanarity and wide-field effects may be significant at low frequencies. Thus, it may be necessary to deal with non-coplanarity and wide fields effects in EPIC. EPIC's antenna-based gridding convolution naturally allows for the non-coplanar effects to be fully incorporated and corrected for \citep{morales_enabling_2011,cornwell_noncoplanar_2008}. A forthcoming paper will elucidate the principles and practicalities behind doing this on EPIC-based imagers.
\item Optimisation - EPIC has unique computational challenges associated with it, which will benefit from broad optimisation of key kernels to remove bottlenecks during the convolutional gridding and the FFT stages.
\item Real-time transient detection - Our transient detection in this manuscript was done using offline analysis of the data. A prototype transient detector has been implemented, however it is in the early stages. An online automated transient detector, with effective RFI filtering, will provide another strong science capability to the EPIC architecture.
\end{itemize}

The addition of aforementioned features will further increase EPIC's scientific repertoire, through correction of antenna based terms in the imaging process, as well as precision calibration in real-time, yielding precision astronomical observations across the whole sky, with high resolution. These are planned to be implemented in the next iteration of development of EPIC on the LWA.  We plan to continue observing in EPIC `mode' at LWA-SV for longer periods of time at a high time resolution, to facilitate blind, source-agnostic surveys where transient phenomena might appear. 


\section*{Acknowledgements}
This work is supported by NSF awards AST-1710719 and AST-1711164.
We gratefully acknowledge the support of NVIDIA Corporation with the donation of a Titan X GPU used for prototyping and testing the EPIC pipeline. Construction of the LWA has been supported by the Office of Naval Research under Contract N00014-07-C-0147 and by the AFOSR. Support for operations and continuing development of the LWA1 is provided by the Air Force Research Laboratory and the National Science Foundation under grants AST-1835400 and AGS-1708855.  A.P.B. is supported by an NSF Astronomy and Astrophysics Postdoctoral Fellowship under award AST-1701440.  J. Kent is funded by the Engineering and Physical Sciences Research Council, UK. We also acknowledge Emma Maton for her help in proof-reading the final manuscript.

\bibliographystyle{mnras}
\bibliography{lwa_epic}{}

\bsp	
\label{lastpage}
\end{document}